\documentclass[epj]{svjour}
\usepackage{graphics}
\usepackage{amsmath}
\begin{document}
\title{Large deviations in spin-glass ground-state energies}
\author{A. Andreanov\inst{1} \and F. Barbieri\inst{1}
\and O.C. Martin\inst{1}
}                     
%
%
\institute{Laboratoire de Physique Th\'eorique et Mod\`eles Statistiques,
b\^at. 100, Universit\'e Paris-Sud, F-91405 Orsay, France}
%
%
\abstract{
The ground-state energy $E_0$ of a spin glass is an example
of an extreme statistic. We consider the large deviations
of this energy for a variety of models when the number
of spins $N$ goes to infinity. In most cases, 
the behavior can be understood qualitatively, in particular
with the help of semi-analytical results for hierarchical lattices. 
Particular attention is paid to the
Sherrington-Kirkpatrick model; after comparing to
the Tracy-Widom distribution which follows from
the spherical approximation, we 
find that the large deviations
give rise to non-trivial scaling laws with $N$.
} 
\maketitle
\section{Introduction}
\label{intro}
Systems with quenched disorder have been studied intensively for
the last two decades. Thermodynamic properties such as the
free energy in these systems fluctuate from sample to sample,
but not very much: indeed, they are self-averaging if
the disorder does not have long range correlations~\cite{LifshitzGredeskul88}.
This means that {\em typical} values of the
free energy density (to name but one quantity) deviate arbitrarily 
little from a fixed value in the large volume limit.
Because of this, little work has considered 
{\em large deviations}, i.e., the probability of finding a 
rare sample (realization of the disorder) for which
the free energy density deviates from the typical value by a significant
amount. 

In spin glasses, the archetype of disordered systems
with frustration, the large deviations of thermodynamic quantities
have been studied so far in only 3 very special cases:
(1) the Random Energy Model~\cite{Derrida80} (REM) which
can be treated completely because of its extreme simplicity;
(2) the Generalized REM~\cite{Derrida85} for which bounds
have been obtained~\cite{BovierKurkova03} on the probability
of large deviations; (3) the 
Sherrington-Kirkpatrick model~\cite{SherringtonKirkpatrick75}
but with results~\cite{Talagrand00} in the paramagnetic phase only.
This last model is important because, in contrast to the 
first two cases, it is based on a microscopic Hamiltonian
with spins, a necessary step to have a realistic model.
In the study that follows, we work well in the spin-glass 
phase for a variety of realistic models,
focusing on the statistics of the system's ground-state 
energy $E_0$. In particular, we consider a nearly
soluble case associated with hierarchical lattices
that gives some qualitative insights
into the properties of the large deviations function.

The outline of this paper is as follows.
We first introduce some definitions and notation
in Section~\ref{sec:gen_results},
recalling some general results on large deviations.
In the rest of the paper,
we tackle successively different models of spin glasses.
Section~\ref{sec:MK} covers a class of hierarchical
models for which both analytic and numerics
can be pushed very far. Section~\ref{sec:REM}
deals with the REM which is exactly solvable.
These two models of spin glasses display 
the importance of summing/minimizing over random variables.
In Section~\ref{sec:SK} the properties of $E_0$ in the 
Sherrington-Kirkpatrick
model are examined numerically;
first its {\em distribution} is compared to the Tracy-Widom
one, and then we investigate the scaling variables entering
into different large deviations functions. In Section~\ref{sec:other_models}
we consider both Edwards-Anderson and mean field
models of spin glasses for which the connectivity
is finite. Our numerical analysis shows that their properties
are close to those found in the hierarchical lattices. Overall
conclusions are given in Section~\ref{sec:conclusions}.

\section{General results on large deviations}
\label{sec:gen_results}

Consider a physical observable $z_N$ of a system
of $N$ spins in the presence of quenched disorder; in our study
$z_N$ will be the ground-state energy density, 
$E_0/N$. Usually, $z_N$ satisfies two convergence properties.
First, the distribution of $z_N$ becomes peaked around
$z^*$ at large $N$:
\begin{gather}
\label{eq:weak_convergence}
\forall \varepsilon >0, \quad P\{|z_N-z^*|\ge \varepsilon \}\to 0
\quad\mbox{as } \  N\to\infty \ .
\end{gather}
A familiar context where this arises is when
averaging a large number of independent
identically distributed (i.i.d. hereafter) random variables:
the weak law of large numbers says that such an
average becomes peaked~\cite{Feller50} when $N \to \infty$.
Second, the {\em sequence} $\{ z_N \}$ itself converges
``almost always'': if one successively adds spins,
thereby increasing $N$ and adding the corresponding
coupling terms to the Hamiltonian, typical sequences
will {\em converge} to $z^*$. In the context of 
averages of i.i.d. random variables,
this is called the strong law of large numbers.
(Note that within Eq.~\ref{eq:weak_convergence},
the sequence $\{ z_N \}$ can deviate arbitrarily from
$z^*$ as long as such deviations arise more and more
rarely as $N \to \infty$.) In physics, the terminology
``self-averaging'' usually refers to the weak convergence:
if $p_N$ is the probability density of $z_N$, 
then $p_N$ converges to the Dirac distribution
$\delta(z_N-z^*)$. It is of interest to understand the
nature of this convergence for physical observables;
often, it turns out to 
be ``exponential''. More precisely, one says
that the distribution of $z_N$ satisfies 
the Large Deviations Principle (LDP in what follows) 
if for all $t$ one 
has the following large $N$ asymptotics:
\begin{gather}
\label{gen_results-ldp}
p_N(z_N=t)\sim K(t,N) \ e^{-N f(t)}
\end{gather}
where $K$ is a slowly varying function compared 
to the exponential. The function $f$ is called the 
Large Deviations Function (LDF hereafter).
In our numerical study, we consider
\begin{gather}
\label{gen_results-estim}
f_N(t) = - \frac{1}{N} \ln \left[ p_N(z_N=t) \right]
\end{gather}
and then we estimate
$f$ via $f_N \to f$ as $N\to\infty$.
More traditionally, the LDF is defined~\cite{Bucklew90} from
the distribution function of $z_N$ which is generally
a better quantity to consider mathematically. Introduce the two
probabilities $P\{z_N\ge t\}$ for $t>z^*$ and $P\{z_N \le t\}$ for $t<z^*$;
from these one defines
the LDF $f$ from the limits (if they exist):
\begin{gather}
\label{gen_results-limit}
f(t)=\lim\limits_{N\to\infty}-\frac{1}{N}\ln P\{z_N\ge t\}\quad t>z^* \notag \\
f(t)=\lim\limits_{N\to\infty}-\frac{1}{N}\ln P\{z_N\le t\}\quad t<z^* \ .
\end{gather}
In the next paragraphs we consider simple examples of the LDP.

\subsection{Sums of independent variables}
\label{sec:sum_indep_var}

Consider first the case of averages of i.i.d. random variables.
Let $\{ x_i\}$ be i.i.d. variables of probability density $\mu$, and
\begin{gather}
z_N = \frac{1}{N} \sum_{i=1}^N x_i \ .
\end{gather}
For any $\lambda$, define
\begin{gather}
{\cal L}(\lambda)= \ln \int \mu(x) e^{\lambda x} dx
\end{gather}
and
\begin{gather}
f(t)=\max\limits_{\lambda}(\lambda t- {\cal L}(\lambda)) \ .
\end{gather}
Cramer's theorem 
(see for instance ref.~\cite{Bucklew90})
states that for any closed set $F$ and any open set $G$
\begin{gather}
\varlimsup_{N\to\infty}\frac{1}{N}\ln P\{z_N\in F\}\le 
-\inf\limits_{t\in F} f(t) \notag \\
\varliminf_{N\to\infty}\frac{1}{N}\ln P\{z_N\in G\}\ge 
-\inf\limits_{t\in G} f(t)
\end{gather}
where $\varliminf$ and $\varlimsup$ are the inf and sup limits.

Still sharper asymptotics are provided by the
Bahadur-Rao theorem if the first two moments of the $x_i$ are finite.
Without loss of generality, assume that the
$x_i$ have a zero mean and a unit variance. Let
$\lambda(t)$ be the value of $\lambda$ such
that the sup in
Cramer's theorem is reached, i.e., 
$f(t)=\lambda t-{\cal L}(\lambda)$.
Then for all $t>0$, we have
\begin{gather}
\lim\limits_{N\to\infty} P\{z_N\ge t\} N^{1/2} e^{Nf(t)} =
{\left(\lambda(t)\sqrt{2\pi {\cal L}^{''}\left[\lambda(t)\right] }\right)}^{-1}
\end{gather}
as well as the analogous relation for $t<0$. 

Let us now apply this framework to one-dimensional spin glasses.
Suppose one has a chain 
of $N$ Ising spins ($S_i = \pm 1$)
with free boundary conditions described by the Hamiltonian
\begin{gather}
H(\{S_i\}) = - \sum_{i=1}^{N-1} J_{i,i+1} S_i S_{i+1}
\end{gather}
where the $J_{i,i+1}$ are i.i.d. random variables
with probability density $\rho$.
It is easy to see that the ground-state energy density is
\begin{gather}
\label{eq:e_0_1d}
e_0=-\frac{\sum\limits_{i=1}^{N-1} |J_{i,i+1}|}{N}
\end{gather}
leading one to the identification $x_i = |J_{i,i+1}|$
and then $e_0=(1-1/N) z_{N-1}$.
The direct application of Cramer's theorem gives
the LDF of the ground-state energy density:
\begin{gather}
f(t)=\max\limits_{\lambda}\left[\lambda t-\ln\int 
\rho(J) ~ e^{-\lambda |J|} dJ \right] \ .
\end{gather}
This last result shows explicitly the non-universality 
of the large deviations function.
To illustrate this, we can compute $f$ for several specific cases.
Consider first the discrete distribution
\begin{gather}
\label{eq:discrete_J}
\rho(J) = \frac{\delta(J)}{2} + \frac{\delta(|J|-1)}{4} \ .
\end{gather}
This model's ground-state 
energy density is always negative and its mean
is $-1/2$. A simple computation shows that
when $t<0$ one has $\exp(\lambda) = (1+t)/(-t)$ while $\lambda=\infty$
for $t \ge 0$. This leads to
\begin{gather}
f(t) = \ln 2 + |t| \ln |t| + (1+t) \ln (1+t)  
\end{gather}
for $-1\le t\le 0$
and $f(t) = \infty$ otherwise. Note that 
$f$ is symmetric about $-1/2$ and that
$f(-1/2)=0$ as it should.
It is easy to see that any discrete $J$ distribution
leads to a bounded range
of possible values of $e_0$; clearly one has
$f(t)=\infty$ outside of this range, but interestingly $f(t)$
is {\em finite} at the boundary values. Finally, for the case considered
in Eq.~\ref{eq:discrete_J}, we have
$f(0)=f(-1)=\ln 2$ but the {\em slope} of $f$ is infinite 
at those points.

Consider now an exponential distribution:
\begin{gather}
\rho(J) = \frac{\alpha e^{-\alpha |J|}}{2} \ .
\end{gather}
The mean (typical) value of $e_0$ is $-1/\alpha$, and again
$e_0\le 0$ but now
{\em all} negative values of $e_0$ can arise. A simple calculation
leads to
\begin{gather}
f(t) = -\alpha t - 1 - \ln ( \alpha |t|) \  \mbox{ for } \ t< 0
\end{gather}
and $f(t)=\infty$ for $t \ge 0$. We now have a logarithmic
divergence as $t \to 0^-$ while the
behavior when $t \to -\infty$ follows that of $-\ln \rho(-|J|=t)$,
i.e., $f(t) \sim - \alpha t$. This is a general feature:
when $t \to -\infty$ in {\em any} spin-glass model, we must
have that all the $J$s become large; furthermore there
is no more frustration, that is we reach the ferromagnetic limit.
Let's consider this explicitly in the case where
$\rho(J)$ is a Gaussian of zero mean and unit variance.
A simple computation shows that
$f(t) \approx t^2/2 - \ln 2$ at large
negative $t$ which has the same divergence
as $-\ln \rho(-|J|=t)$; this pattern will hold in all spin-glass
models. The general picture is then that 
$f(t)=\infty$ for $t>0$ while the $t \to -\infty$ limit simply
reflects the asymptotics of $\rho(J)$. This last
property shows explicitly why
large deviations in spin glasses are not universal.
Note that this is in sharp contrast to what happens for the 
{\em limiting shape} of the distribution of $E_0$! In the case we 
are considering, $E_0$ is the sum of independent variables;
the central limit theorem then shows that 
the distribution's shape becomes Gaussian at large $N$.

\subsection{Minimum of independent variables}
\label{sec:indep_var}

Another quantity of interest in spin glasses is the
domain wall energy $\Delta$. This energy is simply the change in the
ground-state energy when applying anti-periodic 
rather than periodic boundary conditions. 
Using the notation of the previous paragraphs, it is clear
that for a chain of $N$ spins
\begin{gather}
\label{indep_var-dw}
{\Delta}_{N}={\rm sign}(\prod_i J_{i,i+1}) \, \min_{i} \{|J_{i,i+1}|\} \ .
\end{gather} 
The distribution of $\Delta$ is
easily obtained in terms of
$\rho$, especially if $\rho(J)$ is even in $J$. Limiting ourselves
to that case, introduce an integrated distribution function
$q_N(t)$ of ${|\Delta}_N|$:
\begin{gather}
q_N(t)\equiv 2 \int_{|t|}^{\infty} p_N(\Delta_N) d \Delta_N =
q_1^{N-1}(t) \quad \mbox{for} \quad  t\ge 0 \ .
\end{gather}
The LDF $f(t)$ is of course
even and can be obtained directly from 
Eqs.~\ref{gen_results-limit} or \ref{gen_results-estim},
leading to
\begin{gather}
\label{indep_var-ldf-dw}
f(t)=\ln \left[ 2 \int_{|t|}^\infty \rho(J) dJ \right] \ .
\end{gather}
Here again one sees that the LDF is sensitive to $\rho(J)$
and thus is not universal. When we consider instead
the large $N$ shape of the distribution of $|\Delta_N|$, 
we see that it follows a Weibull~\cite{Gumbel58} distribution
if $\rho(J)$ is smooth; such a limit law is universal...

\subsection{Case of general spin-glass models}
\label{sec:dep_var}

For an arbitrary Ising spin glass, the Hamiltonian is
\begin{gather}
\label{eq:H_SG}
H(\{S_i\}) = - \sum_{i<j} J_{ij} S_i S_j
\end{gather}
where the $J_{ij}$ are i.i.d. random variables.
The ground-state energy $E_0$ is the minimum 
value of $H$ when considering
all the $2^N$ assignments of the spin values; the main difference
with the domain wall energy case 
just treated is that these $2^N$ random
values are {\em correlated}. 
It has been proven
for at least some classes of spin glasses~\cite{LifshitzGredeskul88} that
$E_0/N$ converges in probability when $N\to\infty$ ($E_0$
is self-averaging). Our interest in this work
is to find out empirically 
whether this convergence is exponential or not, i.e.,
whether there is a LDP.
It is difficult to motivate a LDP from 
the point of view of minimization over
$2^N$ highly correlated assignments; instead it is better
to formulate the problem differently as follows. Rewrite
Eq.~\ref{eq:H_SG} as a sum of $N$ terms:
\begin{gather}
H(\{S_i\}) = \sum_i \frac{ ( - \sum_{j\ne i} J_{ij} S_j ) }{2} S_j \ .
\end{gather}
The ground-state energy $E_0$ is then the sum of these $N$ terms
when the $S_i$ are set to their ground-state values.
These different terms are {\em correlated} as is clear from
the fact that a given $S_i$ appears in 
several such terms, and in contrast to the
one dimensional chain, there is no way to change variables
so that the terms become independent. Nevertheless,
correlations are relatively weak 
if $H$ is local (that is if the interactions are short range), 
and probably weak correlations do not spoil the
existence of a LDP. In fact, clues to this effect come
from extensions of Cram\'er's theorem, leading us to try to
test for the presence of a LDP in our systems. We now do this
in a ``hands-on'' fashion, using numerical analysis on
different spin-glasses models. 

\section{Migdal-Kadanoff lattices}
\label{sec:MK}

\begin{figure}
\resizebox{0.48\textwidth}{!}{%
  \includegraphics{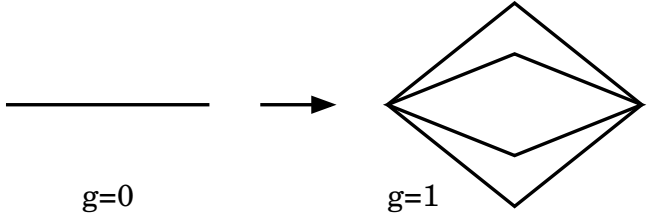}
}
\caption{Construction of a Migdal-Kadanoff lattice having 
$b=4$ branches and $l=2$ segments.}
\label{fig:mk}       
\end{figure}

We begin with the family of models following from the Migdal-Kadanoff (MK) 
approach \cite{SouthernYoung77,BerkerOstlund79} where one 
performs a bond-moving real-space renormalisation group. This 
procedure effectively amounts to computing quantities on hierarchical 
(MK) lattices defined recursively. The recursion takes one bond 
(that is an edge of the current graph) into $b$ paths, each 
made of $l$ segments. The first iteration is represented 
in Fig.~\ref{fig:mk} for $b=4$, $l=2$. In what follows we restrict 
ourselves to the case $l=2$.
%
Given such a ``lattice'', Ising spins are placed
on its sites. The system's Hamiltonian is then
\begin{gather}
\label{eq:HEA}
H_J = - \sum_{\langle ij \rangle} J_{ij} S_i S_j
\end{gather}
where $S_i=\pm 1$ and $\langle ij \rangle$
restricts the sum to nearest neighbors. The $J_{ij}$
are quenched i.i.d. random variables assigned
to the edges. In general, they are either
Gaussian or bimodal ($J_{ij}=\pm1$).

Compared to the 
systems we will consider further on, 
MK lattices have the advantage of allowing an 
exact recursion for the ground-state energy. Thus 
both analytic and numerical computations 
of the LDF can be performed without too much difficulty,
even though $E_0$ remains a sum of {\em dependent} 
random variables.

\subsection{Definitions}
\label{sec:MK_definitions}
If $g$ is the recursion level (beginning with $g=0$
as shown in Fig.~\ref{fig:mk}), then 
the ``linear'' lattice size (or diameter) $L_g$ is $2^g$ and 
the volume $N_g$ (actually the number of edges or terms 
contributing to the energy) is ${(2b)}^g$. The space dimension $d$
is obtained via the identity $N_g=L_g^d$ so one has 
$d=1+\ln b/\ln 2$. The usual choice for $d=3$ is $b=4,l=2$, while 
$b=l=2$ gives $d=2$.

In the recursion relation for the ground-state 
energy, one needs to keep track of two energies: $E^{(p)}_g$ and $E^{(a)}_g$.
These give the MK lattice ground-state energy at the $g$th 
generation when the two exterior spins are respectively 
parallel and antiparallel. The ground-state energy $E_0$ then reads:
\begin{gather}
E_0=\min\{E^{(p)},E^{(a)}\} \ .
\end{gather}
The recursion for a $b$-branch MK lattice with $l=2$ at 
the $(g+1)$th stage is:
\begin{gather}
\begin{split}
\label{MK-main-E-ap}
E^{(p)}_{g+1}=\sum\limits_{k=1}^b \min\{&E^{(p)}_g(1,k)+E^{(p)}_g(2,k),\\
&E^{(a)}_g(1,k)+E^{(a)}_g(2,k)\}
\end{split}\\
\begin{split}
E^{(a)}_{g+1}=\sum\limits_{k=1}^b \min\{&E^{(a)}_g(1,k)+E^{(p)}_g(2,k),\\
&E^{(p)}_g(1,k)+E^{(a)}_g(2,k)\}
\end{split}
\end{gather}
where the index $1,2$ refers to the two bonds forming the $k$th 
branch. These equations say that for each of the $b$ branches, one 
has to choose the orientation of the middle spin
such that the energy is minimized given the orientations
of the external spins. Note that the terms summed are independent 
as they live on different branches. 

The analysis of these recursions requires the study of the joint 
distribution of $E^{(p)}$ and
$E^{(a)}$ which is a complicated 
problem. It simplifies a bit if one takes 
$\Delta=E^{(a)}-E^{(p)}$ and $\Sigma=E^{(p)}+E^{(a)}$ as the new 
variables: this gives an autonomous equation for $\Delta$:
\begin{gather}
\begin{split}
&{\Delta}_{g+1}=\sum\limits_{k=1}^b\\
&\mbox{sign} \Big[ {\Delta}_g(1,k){\Delta}_g(2,k) \Big]
\min\{|{\Delta}_g(1,k)|,|{\Delta}_g(2,k)|\}
\end{split}\\
\begin{split}
\label{MK-main-ds}
&{\Sigma}_{g+1}=\sum\limits_{k=1}^b\\
&\Big[ {\Sigma}_g(1,k)+{\Sigma}_g(2,k) -
\max\{|{\Delta}_g(1,k)|,|{\Delta}_g(2,k)|\} \Big] \ .
\end{split}
\end{gather}
For our work, we consider distributions of the $J_{ij}$ that are 
symmetric about $0$; that of $\Delta$ is then also symmetric about $0$.

\subsection{Domain wall energies}
\label{sec:dw_energies}

\begin{figure}
\resizebox{0.48\textwidth}{!}{%
	\includegraphics{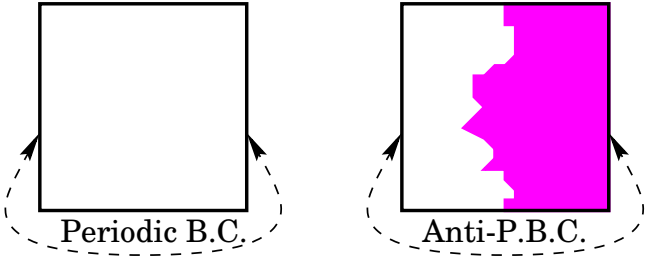}
}
\caption{Domain-wall for the ferromagnetic case: the 
shaded region represents the spins that have flipped.}
\label{fig:fm-dw}
\end{figure}

The $\Delta$ introduced in Eq.~\ref{MK-main-ds} gives the domain wall 
energy for the given sample. The meaning of this quantity 
is best understood through 
the ferromagnetic case presented in Fig.~\ref{fig:fm-dw}:
it is the smallest energy (in absolute value) of domain
walls separating the sample into left and right. The domain 
wall energy is a central quantity in the theory of spin glasses: if the 
typical value of $\Delta$ diverges with $L$, one has a true 
spin-glass phase~\cite{BrayMoore86,FisherHuse86}. 

Interestingly, $\Delta$ is {\em not} self averaging, in contrast
to $E_0$. Nevertheless, one may still consider the large deviations
of this quantity. First, notice that the number of terms
that contribute to $\Delta$ is $O(L^{d-1})$ in MK lattices;
the ferromagnetic case thus leads to $\Delta = O(L^{d-1})$.
This suggests we consider the scaling 
${\Delta}_g\simeq L_g^{d-1}$ when $g\to\infty$. We thus define 
$x_g$ as ${\Delta}_g/L_g^{d-1}$ and hope for a LDP of the type
\begin{gather}
\label{MK-dw-ldp}
p(x_g=t) \simeq e^{-N_g f(t)} \ .
\end{gather}
Let's rewrite the autonomous Eq.~\ref{MK-main-ds} for $x_g$ as:
\begin{gather}
\label{MK-main-dw-n}
\begin{split}
&x_{g+1}=\frac{1}{b}\sum\limits_{k=1}^b\\
&\mbox{sign} \Big[ x_g(1,k)x_g(2,k) \Big]
\min\{|x_g(1,k)|,|x_g(2,k)|\} \ .
\end{split}
\end{gather}
Unlike the case in one dimension, it is impossible 
to apply the technique of 
distribution functions because of the sum over $k$ terms; similarly,
a Fourier transformation is useless because of 
the minimization. Nevertheless, some progress can
be made by noting that the right side 
of Eq.~\ref{MK-main-dw-n} is a partial mean 
of $x_g(\alpha,k)$, $\alpha = 1,2$. Then one can show
that if $x_g$ satisfies the LDP with 
LDF $f(t)$, then so does $x_{g+1}$: once a LDP is 
formed, it is conserved. Empirically, the LDP turns out to be exactly 
as expected in Eq.~\ref{MK-dw-ldp}. Thus, we believe a limit 
$f(t)=-\lim N_g^{-1} \ln p_g(x_g=t)$ as $N_g\to\infty$ 
exists. Clearly,
the LDF isn't universal; it depends on the initial distribution 
$p_0$ of $x_0$.

To obtain an analytical estimate of $f(t)$,
we notice that as $b \to \infty$
the LDP should be 
formed after just one generation. An {\em approximation}
$f_a(t)$ for the LDF $f(t)$ is then
\begin{gather}
\label{MK-dw-ldf}
f_a(t)=\frac{1}{2}\max\limits_{\lambda}\left\{
\lambda t-\ln \Big[ 4\int ds e^{s\lambda}p_0(s)
\int\limits_{|s|}^{\infty}dv p_0(v) \Big] \right\}
\end{gather}
where the argument of the logarithm 
is the characteristic function of 
$\mbox{sign}(x_0(1)x_0(2))\min\{|x_0(1)|,|x_0(2)|\}$.
This formula arises from the direct application of Cram\'er's theorem 
for Eq.~\ref{MK-main-dw-n} with $g=0$,
regarding the quantities 
$\mbox{sign}(x_0(1)x_0(2))\min\{|x_0(1)|,|x_0(2)|\}$ as 
the i.i.d. variables (recall that they live on different branches).

How good is this approximation, and in particular
does it become exact at large $b$? Note that the 
LDF (\ref{MK-dw-ldf}) is $b$-independent because
we took $b\to\infty$ and we expect in fact
$f_a(t)$ to be the limiting value of $f(t)$ as $b \to \infty$.
A series of numerical simulations were carried out in order to 
quantify the large $b$ approximation. We used 
the bimodal distribution $J_{ij}=\pm 1$:
\begin{gather}
p_0(J)=\frac{1}{2}(\delta(J-1)+\delta(J+1)) \ .
\end{gather}
The evaluation of Eq.~\ref{MK-dw-ldf} for $|t|\le2$ then gives:
\begin{gather}
f_a(t)=\frac{1}{2}\left[ -\ln 2+\ln\sqrt{4-t^2}
-\frac{t}{4}\ln \left( \frac{2-t}{2+t} \right) \right]
\end{gather}
which is even as it must; one also has $f_a(t)=+\infty$
for $|t|>2$ which is good since it also holds necessarily
for the exact $f(t)$. 
\begin{figure}
\resizebox{0.48\textwidth}{!}{%
	\includegraphics{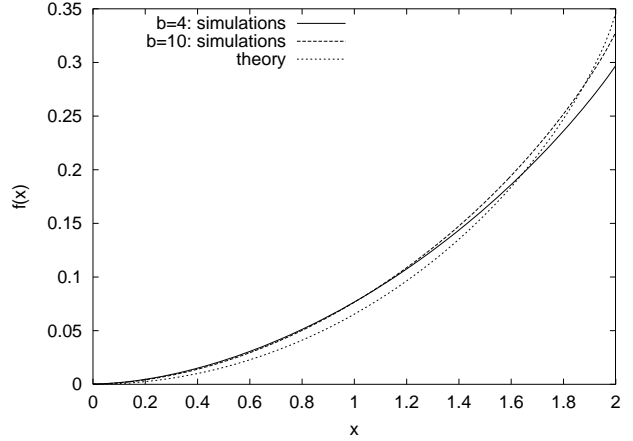}
}
\caption{Domain-wall LDF for the Migdal-Kadanoff lattices
when $J_{ij}=\pm 1$. Displayed are
data for $b=4$, $b=10$, and the theoretical $b=\infty$ 
prediction.}
\label{fig:mk-dw}
\end{figure}

\begin{figure}
\resizebox{0.48\textwidth}{!}{%
	\includegraphics{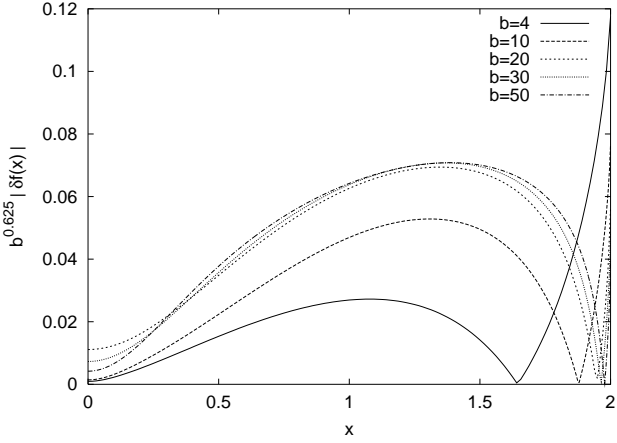}
}
\caption{Rescaled difference between the $b$ finite 
and $b=\infty$ LDF for domain-wall energies within 
the $J_{ij}=\pm 1$ MK model.}
\label{fig:mk-dw-error}
\end{figure}

The simulations revealed a rapid convergence 
with generations of the large 
deviations functions to their limits: $g=5$ was
enough to nearly reach machine precision 
even when $b=4$. Furthermore, as illustrated in 
Fig.~\ref{fig:mk-dw}, the 
discrepancy $\delta f(t)$ between 
the theoretical prediction $f_a(t)$ and the numerically
determined LDF decreased with increasing $b$ 
(a few percent for $b=10$ to fractions of a percent for $b=50$). The 
numerical analysis of $\delta f$ suggests that the corrections 
are of order $b^{-5/8}$. The behaviour of $b^{0.625}\delta f(t)$ is 
shown in Fig.~\ref{fig:mk-dw-error} for several values of $b$.
Naturally, it would be of interest to understand the origin of
this exponent. It is also interesting to note that the range
of $t$ is bounded, $-2 \le t \le 2$, but that the 
derivative of $f$ at these limit points is finite,
in contrast to what happened in the one-dimensional model.

\subsection{Ground-state energies}
\label{sec:gs_energies}

The study of $E_0$ is more complicated as it has no
autonomous recursion equation: we must follow the joint distribution
of $\Sigma$ and $\Delta$.
Recall that
\begin{gather}
\label{MK-E0-Sigma-Delta}
E_0=(\Sigma-\Delta)/2  \ ;
\end{gather}
since ${\Delta}_g$ scales as $L_g^{d-1}$ whereas 
${\Sigma}_g$ scales as $L_g^d$, the $\Delta$ term
in Eq.~\ref{MK-E0-Sigma-Delta} can be 
neglected in the study of large deviations. Thus, the large 
deviations of $\Sigma$ and of $E_0$ coincide, allowing us to focus
hereafter on $\Sigma$.

We define the intensive variable ${\xi}_g={\Sigma}_g/N_g$ and 
rewrite Eq.~\ref{MK-main-ds} as: 
\begin{gather}
\begin{split}
\label{MK-gs-main-n}
&{\xi}_{g+1}=\frac{1}{b}\sum\limits_{k=1}^b 
\Bigl[\frac{1}{2}({\xi}_g(1,k)+{\xi}_g(2,k))\\
&-\frac{1}{2^{g+1}}\max\{|x_g(1,k)|,|x_g(2,k)|\}\Bigr] \ .
\end{split}
\end{gather}
A LDP is expected as $N_g \to \infty$:
\begin{gather}
\label{MK-gs-ldp}
p(\xi_g=t)\sim\exp(-N_g f(t)) \ .
\end{gather}
Note that $\overline \Sigma$, the disorder average of $\Sigma$, 
has a non-zero mean value; 
it can be written 
in terms of the statistics of the $x$ via iteration
of Eq.~\ref{MK-gs-main-n}:
\begin{gather}
\overline{\xi_{g+1}}=-\sum\limits_{p=0}^g
\frac{1}{2^{p+1}}\overline{\max\{|x_k(1)|,|x_k(2)|\}}
\end{gather}
where the index $p$ labels generations.
This identity leads us to guess that the statistics of
${\xi}$ is determined 
entirely by that of $x$. Indeed, it is possible to write down 
explicitly ${\xi}_{g+1}$ in terms of the $x_p$, $0\le p\le g$:
\begin{gather}
\begin{split}
\label{MK-gs-main-xn}
&{\xi}_{g+1}=\\
&-\frac{1}{2}\sum\limits_{p=0}^g
\frac{2^{-p}}{b{(2b)}^{g-p}}\sum\limits_{m=1}^{b{(2b)}^{g-p}}
\max\{|x_p(1,m)|,|x_p(2,m)|\}
\end{split}
\end{gather}
where the different $x$'s have been regrouped
according to their generation number.
Evidently the terms within the sum over $m$ are independent, 
but the terms for different generations are correlated:
indeed, each $x_{p+1}$ is a sum over a sub-set of the
$x_p$'s. Thus the problem hasn't really become easier: instead of the joint 
distribution of $\xi$ and $x$,
we are obliged to consider the joint distribution 
of the $x_p$ at all $p \le g$.
On the other hand it is reasonable to expect weak 
cross-generation correlations for $b\to\infty$. This suggests the 
approximation of {\em independent generations}: take all the 
terms in Eq.~\ref{MK-gs-main-xn} to be uncorrelated! Such
an approximation 
leads to the estimate $\tilde f(t)$ of the LDF:
\begin{gather}
\label{MK-gs-ldf}
\tilde f(t)=\frac{1}{2}\sum\limits_{p=0}^{\infty}
\frac{1}{N_p}\ln M^{\max}_p\left(\frac{t}{2^{p+1}N_p}\right) \ .
\end{gather}
In this equation,
$M_p^{\max}$ represents the characteristic 
function for $\max\{|x_p(1)|,|x_k(2)|\}$.
To actually compute $\tilde f$, we started with
the numerically
determined distribution of the $x_p$ and performed
the sum in the series as far as possible and then estimated
the remaining part from asymptotics.

\begin{figure}
\resizebox{0.48\textwidth}{!}{%
	\includegraphics{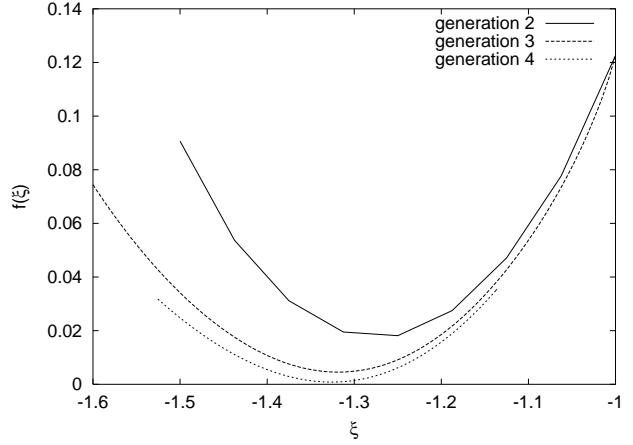}
}
\caption{Convergence of the ground-state 
energy density LDF
with the number of generations 
when $b=4$ in the $J_{ij}=\pm 1$ MK model.}
\label{fig:mk-sg-conv}
\end{figure}

Numerical simulations were carried out to determine
the distributions of $\Sigma$ and of the $x_p$.
As before, we used the bimodal $J_{ij}=\pm 1$ distribution 
of couplings. Although this time it was much harder to 
compute the distribution of $\Sigma$ when $g$
grew, again a good convergence in $g$ was 
found in 5 to 6 generations; this is illustrated
in Fig.~\ref{fig:mk-sg-conv}. Note that since we are dealing with
the $J_{ij}=\pm1$ model, we clearly have
$-2 \le \xi \le 0$ at large $g$, but in fact the upper limit is not
reached for finite $b$.
Also, when $g$ is finite, the ranges
for $\xi$ are slightly different. Finally,
as $g$ grows, it is difficult to follow the
far tail of the distribution of $\Sigma$ because the
probabilities there become extremely small. These different effects
are responsible for the cut-offs
in the curves we show. As another general remark, consider
the ferromagnetic limit. The probability of
having no frustration in the MK lattice is bounded
from below by $2^{-N_g}$; thus necessarily 
$f(-2)$ is finite whereas $f(t)=\infty$ as soon as $t<-2$.

\begin{figure}
\resizebox{0.48\textwidth}{!}{%
	\includegraphics{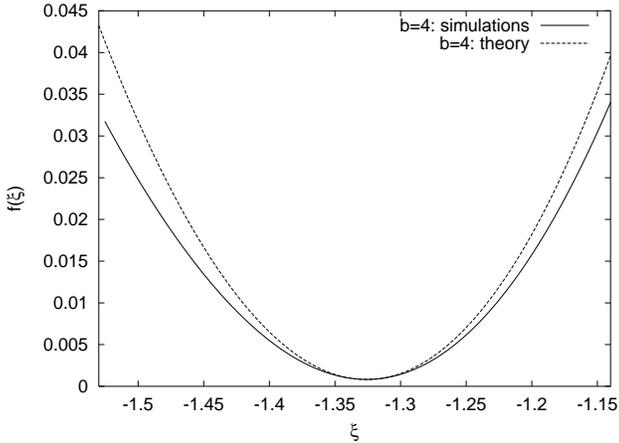}
}
\caption{Ground-state energy density LDF 
when $b=4$ in the $J_{ij}=\pm 1$ MK model: comparison of the independent 
generations approximation (``theory'') and simulations.}
\label{fig:mk-sg-cmp-4}
\end{figure}
How does the estimate of $f$ (obtained from
the data at the largest $g$ we can handle) compare to
$\tilde f$, the prediction of
the independent generations approximation?
\begin{figure}
\resizebox{0.48\textwidth}{!}{%
	\includegraphics{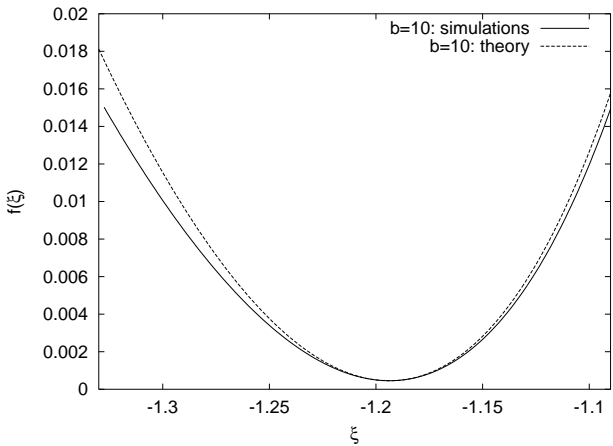}
}
\caption{Ground-state energy density LDF 
when $b=10$ in the $J_{ij}=\pm 1$ MK model: comparison of the independent 
generations approximation (``theory'') and simulations.}
\label{fig:mk-sg-cmp-10}
\end{figure}
The results for $b=4$ and 10 are shown in Figs.~\ref{fig:mk-sg-cmp-4}
and \ref{fig:mk-sg-cmp-10}. The discrepancy 
between ``theory'' and numerics is larger than it was for
$\Delta$ but the error does decrease
slowly as $b\to\infty$, giving evidence
that the independent generations approximation
becomes exact as $b \to \infty$. 

\section{The Random Energy Model}
\label{sec:REM}
The simplest model of spin glasses is
the Random Energy Model (REM)~\cite{Derrida80}. In a system
with $N$ spins there are 
$2^N$ possible assignements of the spins, leading to 
$2^N$ possible energy levels. The framework of the REM is
to consider that these energies are 
independent. The usual choice is for each energy level
to have the probability density
\begin{gather}
p_N(u)=\frac{1}{\sqrt{\pi N}}\exp(-u^2/N)
\end{gather}
whose integrated distribution is
\begin{gather}
q_N(u)=\int_u^{\infty}\frac{dv}{\sqrt{\pi N}}\exp(-v^2/N) \ .
\end{gather}
The ground-state energy $E_0$ for this system is given by:
\begin{gather}
E_0=\min\{x_1,\ldots,x_{2^N}\} \ .
\end{gather}
As such, $E_0$ has follows Gumbel~\cite{Gumbel58} distribution.
However, we are mainly interested in large deviations
that are associated with events where $E_0$ is far from
its typical value. Noting that the $N$ dependence of $p_N$ is chosen
so that $E_0$ scales linearly with $N$ in the large 
$N$ limit, we thus consider $e_0=E_0/N$:
\begin{gather}
\label{eq:e_0}
e_0=\frac{1}{N}\min\{x_1,\ldots,x_{2^N}\} \ .
\end{gather}
Since the $x_i$ are independent, 
the integrated distribution of $e_0$ is:
\begin{gather}
P\{e_0\ge t\} = \Big[ P\{x_1\ge tN\} \Big]^{2^N} = q_N^{2^N}(tN)
\end{gather}
giving
\begin{gather}
p(e_0=t) = N 2^N q_N^{2^N-1}(tN) p_N(tN) \ .
\end{gather}
The asymptotic behavior of these quantities is easily evaluated. For 
$t<-\sqrt{\ln2}$, we have
\begin{gather}
p(e_0=t) \simeq e^{-N(t^2 - \ln 2)}
\end{gather}
in agreement with the fact that the typical value
of $e_0$ is $-\sqrt{\ln2}$.
When $-\sqrt{\ln2} < t < 0$, the nature of the scaling is different :
\begin{gather}
\ln p(e_0=t) \simeq - e^{-N ( t^2 - \ln 2) } \ .
\end{gather}
The scaling at $t<-\sqrt{\ln2}$ can be understood by 
considering the probability that just one of the $2^N$ energies 
is anomalously low. On the contrary, the case $t>-\sqrt{\ln2}$ follows by 
imposing that {\em all} $2^N$ energies are a bit high. We learn from this 
example that the ``normal'' scaling of Eq.~\ref{gen_results-ldp} 
does not always arise, one may have one type of
LDP for $t<\overline{e_0}$ and another 
for $t>\overline{e_0}$.

\section{The Sherrington-Kirkpatrick model}
\label{sec:SK}

The Sherrington-Kirkpatrick~\cite{SherringtonKirkpatrick75} model 
(hereafter SK) is
the mean field limit of spin glasses where all $N$
spins are connected to one another. The Hamiltonian is
\begin{gather}
\label{eq:H_SK}
H_J = - \sum_{i<j} J_{ij} S_i S_j
\end{gather}
the $S_i=\pm 1$ are Ising spins and 
the couplings $J_{ij}$ are Gaussian random variables
of zero mean and variance $1/N$. (This scaling ensures
a good thermodynamic limit, the free energy scaling linearly 
with $N$.) We focus on the statistics of the ground-state energy $E_0$
of this Hamiltonian. For a given sample, $E_0$ can be determined
by combinatorial optimization techniques; we have
used a genetic algorithm~\cite{HoudayerMartin99b}
which finds $E_0$ with a high level
of reliability when $N$ is not too large. This is true
for the SK model and also for the other models we shall
consider further on. In all these cases, we restrict ourselves
to $N$ values where the true $E_0$ is almost certainly obtained.
We then compute the statistics of $E_0$ for a number of different
system sizes $N$ and then attempt to extrapolate to the large 
$N$ limit.

\subsection{Limiting distribution}
\label{sec:SK_distribution}
Before examining the {\em large deviations} of $E_{0}$ in the
SK model, let's first discuss its limiting distribution. The
mean of $E_0$ scales linearly with $N$, and has a variance
that grows, but one may expect the distribution of $E_0$
to have a limiting {\em shape}. To study this, 
one introduces the scaling variable
\begin{gather}
\label{eq:X_J}
X_J = {\frac{E_0 - \overline{E_0} } {\sigma(E_0)} }
\end{gather}
where $\overline{E_0}$ is the disorder average of $E_0$
and $\sigma(E_0)$ its standard deviation.
Bouchaud et al.~\cite{BouchaudKrzakala03} studied the distribution of
$X_{J}$, finding that, for most models of spin glasses it becomes
Gaussian as $N \to \infty$. However 
in the case of the SK, the behavior was clearly non-Gaussian.
Since $E_0$ is an extreme statistic, being the minimum
of $2^N$ energies, it is natural to ask whether 
the distribution of $E_0$ falls into a known
universality class.
There are three standard universality classes for the minimum of  
uncorrelated variables~\cite{Gumbel58}: (1) Gumbel for 
unbounded distributions whose tail decreases faster than 
any power; (2) Fisher-Tippett-Frechet for distributions whose 
tail decreases as a power; (3) Weibull for distributions with cut-offs.
The REM clearly falls into the first class. Interestingly,
the SK does not fall into any of these
three classes~\cite{BouchaudKrzakala03,PalassiniHouches03},
leaving open
the question of the universality class appropriate for its $E_0$.

Recently a new universality class has been uncovered
by Tracy and Widom~\cite{TracyWidom99}: 
in the context of random matrix theory, they found
the limiting distribution of the largest
eigenvalue of an $N \times N$ random matrix when $N \to \infty$.
This distribution is believed to be 
universal, and has already been applied to 
a number of different systems. Note that we are dealing now with the 
maximum of a large number of {\em correlated} random variables;
furthermore, because of the minus sign in  Eq.~\ref{eq:H_SK},
we have to consider in fact $(-1)$ times this largest eigenvalue
when comparing to $E_0$ (note that this changes the sign
of the skewness).
We refer to the distribution of this quantity
within the Gaussian Orthogonal Ensemble as
the Tracy-Widom (TW) distribution. 

\begin{figure}
\resizebox{0.48\textwidth}{!}{%
  \includegraphics{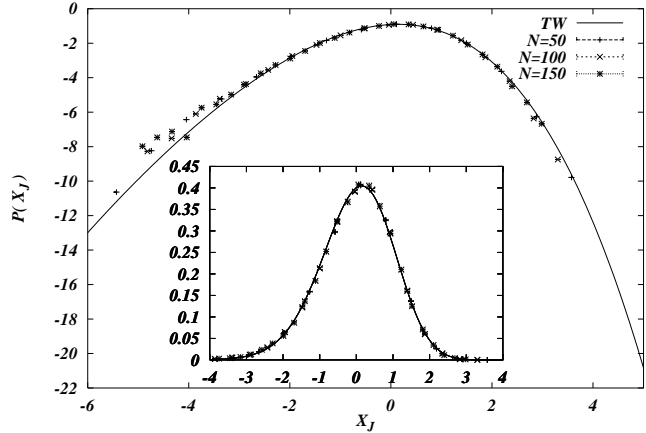}
}
\caption{Comparison of the Tracy-Widom distribution to that
of the Sherrington-Kirkpatrick model at $N=50, 100, 150$.
Data shown are on a logarithmic scale (and on a 
linear scale for the inset); $X_J$ is defined in Eq.~\ref{eq:X_J}.}
\label{fig:sk.TW}       
\end{figure}

How does the distribution of
$E_0$ compare with that of TW? We find that the 
agreement is surprizingly good. In Fig.~\ref{fig:sk.TW} 
we have plotted the TW
distribution as a continuous curve and our
SK data when $N=50, 100, 150$. In the inset we compare the
two distributions directly; the naked eye sees no difference
between the two. In the main part of the figure, we
zoom on the tails, displaying the data
on a logarithmic scale; then definite deviations appear
on the left wing.
At a more quantitative level, we have also compared the
values of skewness $S$ and kurtosis $K$ of the TW and 
SK distributions. Our estimations
are summarized in Table~\ref{tab:sk}.
%
\begin{table}
\caption{Skewness (S) and Kurtosis (K) for the TW distribution 
and for that of $E_0$ in the SK model
(numerical estimates are for $N=50$, $100$, and $150$). }
\label{tab:sk}       
\begin{tabular}{lllll}
\hline\noalign{\smallskip}
$ $ & $TW$ & $SK_{50}$ & $SK_{100}$ & $SK_{150}$ \\
\noalign{\smallskip}\hline\noalign{\smallskip}
$S$ & $-0.293$ & $ -0.43$ & $-0.42$ & $-0.41$ \\
$K$ & $0.165$ & $0.41$ & $0.36$ & $0.39$ \\
\noalign{\smallskip}\hline
\end{tabular}
\end{table}
We find a ``large'' disagreement; clearly the discrepancy seen in the tail of
Fig.~\ref{fig:sk.TW} affects these cumulants, in fact all the more that they
are of higher order. Since at least the SK skewness is numerically stable and
quite different from the TW skewness, we feel that our data indicate that the
two distributions, though very close, are in fact distinct. Thus TW does not
give us the universality class for SK.

On the theoretical side, what possible justification
could be given for $E_0$ to be described by TW
since $E_0$ is a minimum over $2^N$ variables whereas
in the TW matrix problem one considers the extremum of $N$ variables?
One possible answer resides in the 
spherical approximation to the SK model.
Kosterlitz et al.~\cite{KosterlitzThouless76} solved the 
SK model in that approximation and showed that
in the limit $T \to 0$,
the Boltzmann weight becomes dominated by the largest eigenvalue
of the interaction matrix $J_{ij}$; this then leads
to a TW distribution for $E_0$ in that
model. Given this result, it would be good 
to understand why the spherical approximation seems to be
so good for the shape of the distribution of $E_0$ yet
gives a very poor approximation for the ground-state 
energy density ($-1.0$ to be compared with the
actual value of $-0.7633$).

To push further the point that we do not
believe the two distributions to be identical, 
consider the following fact.
If we consider the maximum eigenvalue of a random 
symmetric matrix, the result of TW shows that
the mean shifts with $N$ and the standard deviation 
grows with $N$. Translating into the variable $E_0$,
the correspondence with random matrix theory would predict
that
\begin{gather}
{\overline {E_0}} - N e_0^* \simeq N^{1/3}
\end{gather}
and 
\begin{gather}
{\overline {E_0^2}} -  {\overline {E_0}}^2 \simeq N^{2/3}
\end{gather}
where $e_0^*$ is the thermodynamic limit of ${\overline {E_0} / N}$.
The first equation is in agreement with what happens
in the SK model~\cite{PalassiniThesis,BouchaudKrzakala03},
but the standard deviation does {\em not} at all grow as
$N^{1/3}$ but much more 
slowly~\cite{BouchaudKrzakala03,PalassiniHouches03}.

\subsection{Large deviations}
\label{sec:SK_deviations}

\begin{figure}
\resizebox{0.48\textwidth}{!}{%
\includegraphics{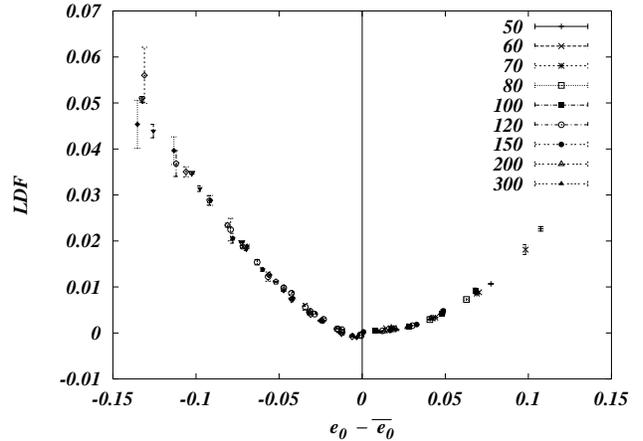} }
\caption{Large Deviations Functions $f_<$ and  
$f_>$ for the Sherrington-Kirkpatrick model
when $N=30$, $50$, \ldots , $300$.}
\label{fig:sk.ld}       
\end{figure}

Still staying within the SK model, we now turn to the large 
deviations of $e_0 = E_0/N$.
The central question is whether there is a LDP.
A naive expectation would be that
\begin{gather}
p_N(e_0) \simeq e^{- N f(e_0)}
\end{gather}
but our data do not follow such a scaling.
By considering the dependence of $\ln p_N(e_0)$
on $N$, our data suggest two quite distinct power
laws in $N$: $N^{1.5}$ when $e_0 > {\overline {e_0}}$
and $N^{1.2}$ when $e_0 < {\overline {e_0}}$.
We thus define two LDF:
\begin{gather}
f_{>}(e_0)=
-\frac{\ln \left[ p_N(e_0) / N^{0.75} \right] }{N^{1.5}} 
\end{gather}
in the first case and
\begin{gather}
f_{<}(e_0)=
-\frac{\ln \left[ p_N(e_0) / N^{0.6} \right] }{N^{1.2}} 
\end{gather}
in the second. The additional rescaling in the
log has no effect on the large $N$ scaling, but
we found it to reduce finite size effects; note that
the motivation for this term comes from Bahardur-Rao 
theorem given in Section~\ref{sec:gen_results}.
We plot in Fig.~\ref{fig:sk.ld}
these two functions modulo a shift of the $x$ axis.

It is not surprizing that the exponent found
for $f_<$ is less or equal to 1.5. Indeed, consider
for instance a sample with $e_0 \le {\overline{e_0}}$.
Now change the sign of $O(N^{1.5})$
$J_{ij}$ that are unsatisfied; this will ``cost''
a probability $O(\exp\left[-A N^{1.5}\right])$. With this change,
we have $e_0 \to e_0 - \delta$ where
$\delta$ is finite, i.e., the desired result. 
The fact that the exponent turns out to be smaller than 1.5
probably comes from the fact that
there are many ways to choose these $O(N^{1.5})$
$J_{ij}$ whereas the argument uses only one
such choice.

Because we are restricted to
$N$ not very large, we cannot exclude 
the possibility that we are seeing effective
exponents and that a different scaling arises
at larger $N$. (In contrast to the MK case,
the large deviations become much more difficult
to measure when $N$ grows and our algorithm
for determining $E_0$ also breaks down at large $N$.)
Nevertheless, since our scalings work well,
the data strongly suggest that there really are 
two different exponents for $e_0 < e_0^*$ and
$e_0 > e_0^*$, just as in the REM.

\subsection{Very large deviations}
\label{sec:SK_VLD}

\begin{figure}
\resizebox{0.48\textwidth}{!}{%
\includegraphics{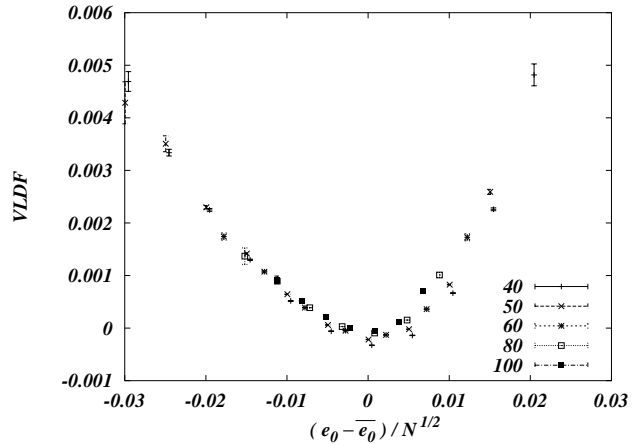} }
\caption{Very Large Deviations Function
for the Sherrington-Kirkpatrick model: data for
$40 \le N \le 100$.}
\label{fig:ferro.sk.ld}       
\end{figure}

In the one-dimensional example of Section~\ref{sec:sum_indep_var}
and in the MK model, the ferromagnetic limit arose when
$e_0 \ll e_0^*$. In the SK model, the absence of frustration
requires constraining $O(N^2)$ $J_{ij}$ and changes the scaling
of $e_0$ which becomes $O(N^{1/2})$. Clearly such
deviations are far rarer than simply having $e_0 \ll e_0^*$.
Because of this, we are limited in our numerical
study to even smaller $N$ values
than before. Nevertheless, 
we have investigated these deviations by considering
the probability distribution of $e_0/N^{1/2}$. We find
\begin{gather}
\label{eq:ferro.sk.ld}
p_N(e_0/N^{1/2}) \simeq N e^{-N^2 f(e_0/N^{1/2})}
\end{gather}
with a very large deviations function (VLDF) $f$
displayed in Fig.~\ref{fig:ferro.sk.ld}. (To
reduce finite $N$ effects, we have in fact used 
$e_0-\overline{e_0}$ rather than $e_0$ itself, but this does
not change the asymptotics.)
We see that for positive arguments, the estimates
for $f$ increase with $N$ which is compatible with
the expectation that $f(t)=\infty$ when
$t>0$. For negative arguments, the data collapse
quite well, justifying the VLDP proposed in
Eq.~\ref{eq:ferro.sk.ld}.

\section{Other spin-glass models models}
\label{sec:other_models}
Most probably, the subtleties of the SK model
arise from the fact that all spins are connected
to one-another. In more realistic models, 
spins interact with just a few neighbors. We thus
return to models of the type described by the
Hamiltonian in Eq.~\ref{eq:HEA}, and now
impose the number of neighbors to be fixed and constant.
The two standard classes of models that do that are
the Edwards-Anderson~\cite{EdwardsAnderson75} model
on square or cubic lattices
and the mean field fixed connectivity 
models~\cite{DominicisGoldschmidt89}.
At present, no analytical
tools are available for treating large deviations, so we
resort to a purely numerical analysis.
In what follows, the disorder variables 
$J_{ij}$ of Eq.~\ref{eq:HEA} are i.i.d. Gaussian 
random variables of zero mean and unit variance.

\subsection{Euclidean models}
\label{sec:Euclidean}

\begin{figure}
\resizebox{0.48\textwidth}{!}{%
\includegraphics{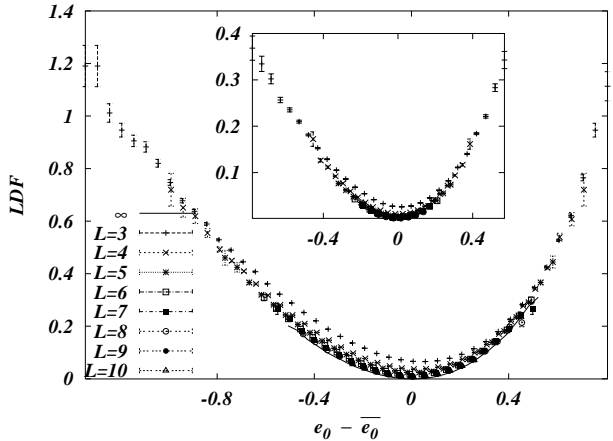} }
\caption{Large Deviations Function in the
Edwards-Anderson model with Gaussian couplings in $d=2$.
The solid line is the limiting curve 
for $N \to \infty$. Inset: the $d=3$ case.}
\label{fig:gea}       
\end{figure}

We first look at the LDF of $e_{0}$ for the 
Edwards-Anderson model 
with Gaussian couplings in two and three dimensions.
In the case of finite dimensional lattices, we know 
that the variance of $E_0$ grows linearly with $N$, and most
probably the distribution of $E_0$ is 
Gaussian~\cite{WehrAizenman90}. This
suggests that the terms contributing to the ground-state 
energy are nearly independent, just as in the MK case.
Then the functions
\begin{gather}
\label{eq:EA}
f_{N}(e_0)=\frac{- \ln \left[ p_N(e_0)/N^{1/2} \right] } {N}
\end{gather}
should converge as $N\to \infty$.
We first checked this on our $3\le L \le 10$ data in 
the two-dimensional ($d=2$) case. Under the hypothesis 
that the finite size corrections are $O(1/N)$, we extract
the limiting function $f_{\infty}$ for $N \to \infty$. This
function is shown 
in Fig.~\ref{fig:gea} as a continuous curve.
The LDF is asymmetric, increasing more rapidly on the 
right than on the left.
We also show our data for the $f_N$, showing that
the scaling in Eq.~\ref{eq:EA} works well.
The $d=3$ case is displayed as an inset; note that
the LDF is more symmetric than for $d=2$.
Also, the convergence of $f_{N}$ to 
$f_{\infty}$ again goes as $O(1/N)$.
   
\subsection{Fixed connectivity random graphs}
\label{sec:FCSG}

We now focus on random graphs that are of
fixed connectivity $z$ at each site. Such graphs 
can be generated
by the following simple algorithm. First
construct $N$ vertices,
then successively add edges to the graph
by randomly connecting sites whose current connectivity
is less than $z$. This process can be made more efficient
by keeping a list of all the candidate sites. The size of this list
decreases; if it reaches $1$, the construction has to be 
abandoned and restarted. In practice such a failure does
not occur very often.

Just as in the Euclidean case, 
we find the simple scaling 
\begin{gather}
p_N(e_0) \simeq A \sqrt{N} e^{- N f(e_0)} \ .
\end{gather}
Thus as in the MK case, we are closer to a sum
of i.i.d. random variables than to a minimum of independent
variables.
\begin{figure}
\resizebox{0.48\textwidth}{!}{%
  \includegraphics{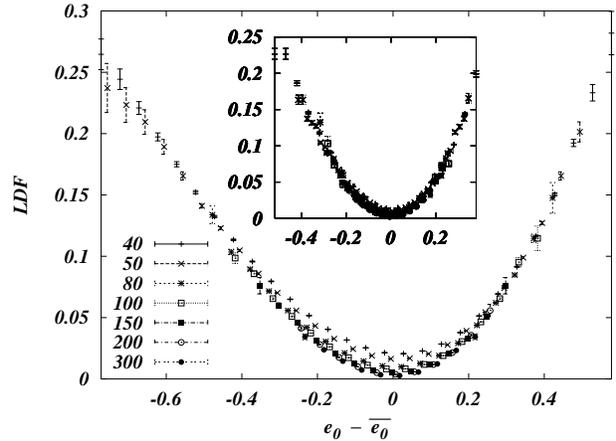}
}
\caption{Large Deviations Function for fixed connectivity random 
graphs when $z=10$. Inset: the case $z=3$.}
\label{fig:fcsg}       
\end{figure}
In Fig.~\ref{fig:fcsg} we display the LDF for the case
$z=10$; the inset is for $z=3$. We see that the
convergence to the large $N$ limit is quite good,
leading to an envelope curve.
It should be clear that the two functions displayed are
distinct: the case $z=3$ is 
a bit more symmetric than the case $z=10$.

\section{Conclusions}
\label{sec:conclusions}
We have investigated the large deviations of the ground-state 
energy $E_0$ in several Ising spin-glass models. There are two 
extreme cases: (1) $E_0$ is the sum of independent random 
variables (as in a one dimensional lattice); (2) $E_0$ is the minimum 
of a large number of independent random variables (as in the 
Random Energy Model). The case of the Sherrington-Kirkpatrick model 
led to a behavior intermediate between these two extremes.
However, the other spin-glass models we considered, 
all of which have finite connectivity, resemble closely the 
first case. This was particularly patent for the hierarchical 
lattice models where high quality numerical computations were 
possible as well as analytical estimates.

We also pointed out that the large deviations function is
not universal; in particular, its tails depends very much on
the far distribution of the disorder variables $J_{ij}$.
This leads one to ask whether there is any universality in 
large deviations functions.
This question is all the more interesting that
Brunet and Derrida~\cite{BrunetDerrida00}
found the large deviations function 
for a directed polymer in a random medium
to be universal; furthermore, that function is the same as the one describing
{\em infinitesimal} deviations in their system. To 
have this kind of universality
in spin glasses, the deviations must not affect much the
values taken by the $J_{ij}$; for us, this seems possible only
in the Sherrington-Kirkpatrick model. More precisely,
we believe that the regime
$E_0 = O(N)$ in that model {\em is} universal (independent of
the details of the $J_{ij}$); this is to be contrasted with
the ferromagnetic limit where clearly the large deviations function is
not universal.

Many questions surrounding these issues are still
very open; let us list a few.
(0) What are the analytical properties of the large deviation
function, and when is it related to the distribution of
infinitesimal deviations?
(1) When the large deviations function is finite, we found
a quite smooth convergence of our finite $N$ estimates to a 
limiting curve; is this convergence uniform?
(2) What is the distribution of the disorder variables
$J_{ij}$ when $e_0$ deviates from its typical value?
(3) In the ferromagnetic limit, do the regions with frustration
phase separate out? (We believe not.)
Clearly, numerical analysis is not a good tool to 
tackle most of these questions. Fortunately, there is good reason
to believe that
Migdal-Kadanoff lattices provide a good framework to address 
these questions; we hope that analytical
results will be derived in the near future for such spin-glass
models.

{\em Acknowledgements --} 
We thank A. Pagnani for
his help throughout this project and C. Tracy
for providing us with his Mathematica program
for generating the TW distribution.
We also thank
O. Bohigas, J.-P. Bouchaud, B. Derrida, E. Marinari,
M. M\'ezard, R. Monasson and G. Parisi for their 
critical comments. 
This work was supported in part by the European Community's
Human Potential Programme (contracts
HPRN-CT-2002-00307 for DYGLAGEMEM and
HPRN-CT-2002-00319 for STIPCO).


\bibliographystyle{prsty}
\bibliography{../../../Bib/references,../../../Bib/largeDev}

\end{document}